\documentclass[12pt]{article}
\usepackage{graphicx}

\newcommand\pubnumber{SNSN-323-63}
\newcommand\pubdate{\today}

\def\napoli{Department of Physics, Tohoku University\\
6-3, Aramaki Aza-Aoba, Aoba-ku, Sendai, Miyagi 980-8578, Japan}
\def\support{}

\def\Title#1{\begin{center} {\Large #1 } \end{center}}
\def\Author#1{\begin{center}{ \sc #1} \end{center}}
\def\Address#1{\begin{center}{ \it #1} \end{center}}

\newcommand\pubblock{\rightline{\begin{tabular}{l} \pubnumber\\
         \pubdate  \end{tabular}}}
\newenvironment{Abstract}{\begin{quotation}  }{\end{quotation}}
\newenvironment{Presented}{\begin{quotation} \begin{center} 
             PRESENTED AT\end{center}\bigskip 
      \begin{center}\begin{large}}{\end{large}\end{center} \end{quotation}}

\def\beq{\begin{equation}}
\def\eeq#1{\label{#1}\end{equation}}
\def\eeqn{\end{equation}}

\def\beqa{\begin{eqnarray}}
\def\eeqa#1{\label{#1}\end{eqnarray}}
\def\eeqan{\end{eqnarray}}

\let\bar=\overbar

\def\Dslash{\not{\hbox{\kern-4pt $D$}}}
\def\dslash{\not{\hbox{\kern-2pt $\del$}}}

\def\msb{{\bar{\ssstyle M \kern -1pt S}}}

\begin{document}
\begin{titlepage}
\pubblock

\vfill
\Title{Belle time-dependent gamma measurements}
\vfill
\Author{ Yoshiyuki Onuki\support}
\Address{\napoli}
\vfill
\begin{Abstract}
The Belle experiment has measured the CKM angle $\gamma$ in a variety of
ways.
In this paper, 
we focused on the recent progress of time-dependent $\gamma$ analysis 
and the related measurements in Belle.
\end{Abstract}
\vfill
\begin{Presented}
6th International Workshop on the CKM Unitary
Triangle, University of Warwick UK, 6-10 September 2010
\end{Presented}
\vfill
\end{titlepage}
\def\thefootnote{\fnsymbol{footnote}}
\setcounter{footnote}{0}

\section{Introduction}

The phenomenon of $CP$ violation that is observed 
in high energy physics experiments is 
explained by a irreducible complex phase 
in the CKM matrix\cite{N.Cabibbo} in the Standard Model. 
The phase can be derived from measurements of the three angles and sides
of the Unitarity Triangle.
The angles are called as
$\alpha = Arg[-(V_{td}V_{Vtb}^{*})/(V_{ud}V_{ub}^{*})]$, 
$\beta = Arg[-(V_{cd}V_{cb}^{*})/(V_{td}V_{tb}^{*})]$ and 
$\gamma = Arg[-(V_{ud}V_{ub}^{*})/(V_{cd}V_{cb}^{*})]$. 


The angle $\gamma$ is the least well determined among all angles.
In the time-dependent $\gamma$ measurement of $D^{(*)}\pi$, 
at first $CP$ violation phase $2\beta$ is appeared in the $B\bar{B}$ 
mixing process and $\gamma$ arised from followed two decay paths: 
Cabibbo favored decay (CFD) and doubly Cabibbo suppressed decay (DCSD) 
into a final state.
The time-dependent decay rate is given by
\begin{eqnarray}
\begin{array}{rll}
\displaystyle
P(B^{0}\rightarrow D^{(*)+}\pi^{-})&=&\frac{1}{8\tau_{B^{0}}}e^{-|\Delta t|/\tau_{B^{0}}}[1-C\cos(\Delta m \Delta t)-S^{+}\sin(\Delta m \Delta t)],\\
P(B^{0}\rightarrow D^{(*)-}\pi^{+})&=&\frac{1}{8\tau_{B^{0}}}e^{-|\Delta t|/\tau_{B^{0}}}[1+C\cos(\Delta m \Delta t)-S^{-}\sin(\Delta m \Delta t)],\\
P(\bar{B}^{0}\rightarrow D^{(*)+}\pi^{-})&=&\frac{1}{8\tau_{B^{0}}}e^{-|\Delta t|/\tau_{B^{0}}}[1+C\cos(\Delta m \Delta t)+S^{+}\sin(\Delta m \Delta t)],\\
P(\bar{B}^{0}\rightarrow D^{(*)-}\pi^{+})&=&\frac{1}{8\tau_{B^{0}}}e^{-|\Delta t|/\tau_{B^{0}}}[1-C\cos(\Delta m \Delta t)+S^{-}\sin(\Delta m \Delta t)],
\end{array}
\end{eqnarray}
where $\Delta t$ is the difference between the decay-time of the signal B  
and other $B$, $\tau_{B^{0}}$ is the average neutral 
$B$ meson lifetime, $\Delta m$ is the $B^{0}-\bar{B^{0}}$ mixing parameter, 
and $C = (1-R^2)/(1+R^2)$. $S^{\pm}$ are given by
\begin{eqnarray}
S^{\pm} = \frac{2(-1)^{L}R\sin(2\beta+\gamma \pm \delta)}{(1+R^{2})},
\end{eqnarray}
where $R$ is the ratio of the magnitudes of the DCSD and CFD, 
$L$ is the orbital angular momentum of the final state(1 for $D^{*}\pi$
and 0 for $D\pi$), and $\delta$ is the strong phase difference of the CFD
and DCSD. 

\section{$D^{(*)}\pi$ time-dependent $CP$ analysis}

The time-dependent $CP$ violation analysis with fully reconstructed 
$D^{(*)}\pi$ events from a data sample of 
$386\times 10^{6} B\bar{B}$ pairs had performed by Belle\cite{Sarangi}. 
The used decays are 
$D^{*+}\rightarrow D^{+}\pi^{0}$ or $D^{0}\pi^{+}$ 
followed $D^{+}\rightarrow K^{-}\pi^{+}\pi^{+}$
and $D^{0}\rightarrow K^{-}\pi^{+}$,$K^{-}\pi^{+}\pi^{0}$,
$K^{-}\pi^{+}\pi^{+}\pi^{-}$ and 
$K^{0}_{S}\pi^{+}\pi^{-}(K^{0}_{S}\rightarrow \pi^{+}\pi^{-})$.
Determination of the flavor of $B$ meson opposite to $CP$ side of $B$ meson
is used leptons, pions and kaons which are not associated with $CP$ side
$B$ meson.
Tag-side interference is taken into accounts introducing a small
asymmetry when daughter particles from hadronic decays such as 
$D^{(*)}\pi$ are used for the flavor tagging, due to the same $CP$ 
violating effect\cite{Long}. 
The results are 
\begin{eqnarray}
\begin{array}{lll}
\displaystyle
S^{+}(D^{*}\pi)&=& 0.050 \pm 0.029 (stat) \pm 0.013 (syst),\\
S^{-}(D^{*}\pi)&=& 0.028 \pm 0.028 (stat) \pm 0.013 (syst),\\
S^{+}(D\pi)    &=& 0.031 \pm 0.030 (stat) \pm 0.012 (syst),\\
S^{-}(D\pi)    &=& 0.068 \pm 0.029 (stat) \pm 0.012 (syst),
\end{array}
\end{eqnarray}
where the errors are statistical and systematic error, respectively.

The time-dependent $CP$ violation analysis with partially reconstructed 
$D^{*}\pi$ from a data sample of 
$657\times 10^{6} B\bar{B}$ pairs had updated by Belle\cite{Bahinipati}. 
The measurement required fast pion($\pi_{f}$) and slow pion ($\pi_{s}$) 
candidates. 
Three kinematic variables, $p_\delta$,$p_\parallel$ and $p_\perp$ 
are defined and the cut are applied to reject backgrounds. 
In the boost the $\pi_{f}$ into the partially reconstructed $D^{*}$ frame,
$p_\parallel$ and $p_\perp$ are defined parallel and the transverse 
components of the momentum of the $\pi_{s}$ along with the opposite 
direction to $\pi_{f}$. The $p_{\delta}$ is defined as 
$p_\delta \equiv |p_{\pi_f}|-|p_{D^{*}}|$, where $|p_{D^{*}}|$ is magnitude
of momentum for $D^{*}$ which reconstructed by energies of $B$ meson and 
$\pi_{f}$.
Three categories of background source are defined: 
$D^{*\mp}\rho^{\pm}$, 
correlated background originated from inclusive $D^{*}$ decay,
uncorrelated background which includes everything else. The fractions 
are determined from $(p_{\delta}, p_{\parallel})$ two dimensional fit.  
The flavor tagging
is used by requiring a high momentum lepton in the event. 
This helps reducing continuum background of 
$e^+e^-\rightarrow q\bar{q}$, where $q=u,d,s$ and $c$.
The result is 
\begin{eqnarray}
\begin{array}{lll}
\displaystyle
S^{+}(D^{*}\pi)&=& 0.057 \pm 0.019 (stat) \pm 0.012 (syst),\\
S^{-}(D^{*}\pi)&=& 0.038 \pm 0.020 (stat) \pm 0.010 (syst).
\end{array}
\end{eqnarray}

\section{$R_{D^{(*)}\pi}$ measurements with $D^{*}\pi^{0}$ and $D^{(*)}_{s}\pi$}

It is difficult to determine $R_{D^{*}}$ from $B^{0}$ decays
because the DCSD amplitude is small compared to the contribution
from mixing followed CFD, 
$B^{0}\rightarrow \bar{B^{0}}\rightarrow D^{*+}\pi^{-}$.
%
%
Using available branching fraction measurements, $R_{D^{*}\pi}$
can be expressed as
\begin{eqnarray}
\displaystyle
R_{D^{*}\pi}&=& \sqrt{\frac{\tau_{B^{0}}}{\tau_{B^{+}}}\frac{2{\cal B}(B^{0}\rightarrow D^{*+}\pi^{0})}{{\cal B}(B^{0}\rightarrow D^{*-}\pi^{+})}}.
\end{eqnarray}
The decay $B^{+}\rightarrow D^{*+}\pi^{0}$ is searched with a data sample of 
$657\times 10^{6} B\bar{B}$ pairs by Belle\cite{Iwabuchi}. 
The obtained branching fraction is
${\cal B}(B^{+}\rightarrow D^{*+}\pi^{0})=[1.2^{+1.1}_{-0.9}(stat)^{+0.3}_{-0.9}(syst)]\times 10^{-6}$.
The upper limit is ${\cal B}(B^{+}\rightarrow D^{*+}\pi^{0})<3.6\times 10^{-6}$ at $90\%$ confidence level.
This result can be used to set an upper limit on the $R_{D^{*}\pi}$,
\begin{eqnarray}
\displaystyle
R_{D^{*}\pi}&<& 0.051 (90 \% C.L.).
\end{eqnarray}

If we assumed SU(3) flavor symmetry, also $R_{D^{(*)}\pi}$ is given by
\begin{eqnarray}
\displaystyle
R_{D^{(*)}\pi}&=& \tan\theta_{C} \frac{f_{D^{(*)}}}{f_{D^{(*)}_{s}}}\sqrt{\frac{{\cal B}(B^{0}\rightarrow D^{(*)+}_{s}\pi^{-})}{{\cal B}(B^{0}\rightarrow D^{(*)-}\pi^{+})}},
\end{eqnarray}
where $\theta_{C}$ is the Cabibbo angle, and  
$f_{D^{(*)}}$ and $f_{D^{(*)}_{s}}$ are the meson decay constants.

The $R_{D\pi}$ using $D_{s}\pi$ is also measured with
a data sample of $657\times 10^{6} B\bar{B}$ pairs by Belle\cite{A.Das}.
The $D_{s}^{+}$ is reconstructed from 
$D_{s}^{+}\rightarrow (K^{+}K^{-})_{\phi}\pi^{+}$,$(K^{-}\pi^{+})_{K^{*}(892)}K$, $(\pi^{+}\pi^{-})_{K_{S}}K^{+}$.
The obtained branching fraction is  
${\cal B}(B^{0}\rightarrow D_{s}^{+}\pi^{-}) = (1.99\pm 0.26(stat) \pm 0.18(syst))\times 10^{-5}$. 
Using the fraction,  Cabibbo angle\cite{N.Nakamura} 
$\tan\theta_{C} = 0.2314\pm 0.0021$, the lattice QCD calculation
of $f_{D_{s}}/f_{D} = 1.164 \pm 0.011$\cite{E.Follana} and the 
fraction 
${\cal B}(B^{0}\rightarrow D^{-}\pi^{+})=(2.68\pm0.13)\times 10^{-3}$\cite{N.Nakamura},
we obtain 
\begin{eqnarray}
\displaystyle
R_{D\pi}&=& (1.71 \pm 0.11(stat) \pm 0.09(syst) \pm 0.02(theo))\%,
\end{eqnarray}
where the last term accounts for the theoretical uncertainty in the 
$f_{D_{s}}/f_{D}$ estimation. 

The $R_{D^{*}\pi}$ is measured using $D^{*}_{s}\pi$ with a data sample of 
$657\times 10^{6} B\bar{B}$ pairs by Belle \cite{N.J.Joshi}.
The $D_{s}^{*+}$ is reconstructed by 
$D_{s}^{+}$  combining $\gamma$ followed 
$D_{s}^{+}\rightarrow (K^{+}K^{-})_{\phi}\pi^{+}$,$(K^{-}\pi^{+})_{K^{*}(892)}K$, $(\pi^{+}\pi^{-})_{K_{S}}K^{+}$.
The yields are extracted from simultaneous fit for above decays. 
The obtained branching fraction is
${\cal B}(B^{0}\rightarrow D_{s}^{*+}\pi) = (1.75 \pm 0.34(stat) \pm 0.17(syst) \pm 0.11({\cal B}))\times 10^{-5}$ 
with significance of $6.1$ standard deviation.
The third error is from uncertainties in the $D_{s}^{+}$ 
decay branching fractions. 
Using the observed fraction, 
${\cal B}(B^{0}\rightarrow D^{*-}\pi^{+})=(2.76\pm 0.13)\times 10^{-3}$, 
$\tan\theta_{C}$\cite{C.Amsler}, 
and the theoretical estimate of the ratio 
$f_{D_{s}}/f_{D}$\cite{E.Follana}, we obtain
\begin{eqnarray}
\displaystyle
R_{D^{*}\pi}&=& (1.58 \pm 0.15(stat) \pm 0.10(syst) \pm 0.03(theo))\%,
\end{eqnarray}
where the third error is theoretical uncertainty in the 
$f_{D_{s}^{+}}/f_{D^{+}}$ estimation.
We have assumed that the ratio $f_{D_{s}}/f_{D} = f_{D_{s}^{*}}/f_{D^{*}}$.
The quenched QCD approximation (heavy quark effective theory) 
predicts\cite{M.Neubert} 
the uncertainty of the assumption about $1\%$, which is included in 
the theoretical uncertainty. 
Uncertainties due to SU(3) symmetry breaking effects\cite{M.A.Baak}, 
which are of order $(10-15)\%$, are not included 
in the theoretical uncertainty above $R_{D\pi}$ and $R_{D^{*}\pi}$ 
with $D_{s}^{(*)}\pi$ decays.

\section{Conclusion}

Time-dependent $\gamma$ analyses in Belle are progressing in Belle.
The results from the partial reconstruction method 
have been updated with a data sample almost twice larger than the previous 
Belle analysis.
Also, the measurements of $R_{D^{(*)}\pi}$ are updated. 

\end{document}